# Promoting Shared Energy Storage Aggregation among High Price-Tolerance Prosumer: An Incentive Deposit and Withdrawal Service

Xin Lu[1], Jing Qiu[1], Cuo Zhang[1], Gang Lei[2], Jianguo Zhu[1]
1. The University of Sydney, Australia
2. University of Technology Sydney, Australia

*Abstract-* **Many residential prosumers exhibit a high price-tolerance for household electricity bills and a low response to price incentives. This is because the household electricity bills are not inherently high, and the potential for saving on electricity bills through participation in conventional Shared Energy Storage (SES) is limited, which diminishes their motivation to actively engage in SES. Additionally, existing SES models often require prosumers to take additional actions, such as optimizing rental capacity and bidding prices, which happen to be capabilities that typical household prosumers do not possess. To incentivize these high price-tolerance residential prosumers to participate in SES, a novel SES aggregation framework is proposed, which does not require prosumers to take additional actions and allows them to maintain existing energy storage patterns. Compared to conventional long-term operation of SES, the proposed framework introduces an additional short-term construction step during which the energy service provider (ESP) acquires control of the energy storage systems (ESS) and offers electricity deposit and withdrawal services (DWS) with dynamic coefficients, enabling prosumers to withdraw more electricity than they deposit without additional actions. Additionally, a matching mechanism is proposed to align prosumers' electricity consumption behaviors with ESP's optimization strategies. Finally, the dynamic coefficients in DWS and trading strategies are optimized by an improved deep reinforcement learning (DRL) algorithm. Case studies are conducted to verify the effectiveness of the proposed SES aggregation framework with DWS and the matching mechanism.**

## I. INTRODUCTION

The rapid growth and widespread utilization of distributed energy systems (DERs), particularly rooftop photovoltaics (PVs), have significantly changed the energy supply worldwide[1]. This proliferation of PV has resulted in an abundance of energy supply during midday, producing surplus energy sold to electricity retailers at a reduced feed-in tariff (FIT). Moreover, the surplus PV generation has led to a shift in peak demand, now occurring in the evening instead of midday [2, 3]. As a result, wholesale electricity prices have also undergone a similar temporal shift [4, 5]. To maximize the benefits of PV installations and reduce household electricity expenses, households with PVs often install energy storage systems (ESS) [6, 7]. The ESS can store the excess PV generation from midday for subsequent utilization during evening hours.

Conventional studies focused on the sizing and economic feasibility of household PV-ESS systems [8]. Cucchiella et al. [9] examined the residential PV and ESS scale in an Italian electricity market. They found the self-consumption levels were crucial for the profitability of PV. However, they also concluded that household ESS did not yield profitability. Similarly, Joern et al. [10] investigated the PV-ESS systems in Germany, considering eight electricity price scenarios from 2013 to 2022. They proposed a cost-optimal investment model and observed that the higher electricity retailer price and the lower wholesale price could enhance the profitability of ESS. Ren et al. [11] analyzed the PV-ESS systems with different electricity price structures, PV and battery sizes in Australia. Their findings indicated that PV-ESS systems did not significantly improve profitability compared to PV-alone systems. Another study conducted by Khalilpour et al. [12] revealed that the benefits of ESS became evident when the FIT was substantially lower than the retail electricity prices. Collectively, the aforementioned studies suggest that despite the efforts in optimizing the capacity of household ESS to match PV, there is no guarantee of enhanced profitability. As a result, a common approach to increase the profitability of individual household ESS involves aggregating them together and implementing new business models to increase profits.

Shared energy storage (SES) [13, 14] emerges as a prominent approach for aggregating individual ESS. Many SES architectures exist [15], including SES being aggregated from multiple household ESS, referred as 'interconnected SES', or independent SES being directly established, referred as 'independent SES'. Wang et al. [16] proposed an interconnected SES facilitating geographically distributed household ESS to trade capacity with others. In [17], households were enabled to share their ESS capacities with large buildings. However, the control of these ESS was managed by different households. Game theory, with subtypes such as the Nash game [16] and noncooperative Stackelberg game [17], was often employed to address the challenges in the above scenarios. However, it is important to note that these game-theoretical approaches often result in stable solutions rather than optimal ones, implying interconnected SES's relatively low utilization efficiency.

On the other hand, independent SES is typically overseen by an independent energy service provider (ESP), responsible for determining the shared capacity price for prosumers. To manage the interactions between SES and prosumers, the ESP of an independent SES must define a market mechanism. Mediwaththe et al. [18] proposed an energy trading system for SES and a pricing incentive plan, with the ESP optimizing the profit by providing price signals and the retailer minimizing the total users' cost. An application of SES among local integrated energy systems was explored in [19], where the authors transformed the model into a two-step optimization problem. The first step was to obtain the optimal scheduling strategy of each energy system, while the second focused on obtaining the optimal pricing information. However, these independent SES often prioritized internal pricing over actively participating in electricity market trading.

The independent ESP of an SES is often engaged in the wholesale market, capitalizing on the opportunity to buy electricity at low prices and sell it at high prices, thereby achieving satisfying profits [15]. A credit-based sharing model for SES was provided in [20]. This approach, along with addressing the ESS dispatch problem, aimed to maximize the profits for both the ESP and the prosumers in the wholesale electricity market. In another study by Li et al. [21], an SES was utilized to offer energy storage rental services, with dynamic rental prices being optimized by the ESP. The ESP also efficiently dispatched the SES to participate in the electricity market. The rental price optimization problem and the ESS scheduling problem were decoupled using a Markov decision process (MDP) and effectively solved using a deep reinforcement learning (DRL) algorithm. It is assumed in these models that users are price-sensitive, meaning any improvement in profit or reduction in electricity bills would incentivize prosumers to respond accordingly. However, it is essential to acknowledge that in real-world scenarios, the responsiveness to renting capacity or energy in the SES would undoubtedly vary from those given by [20] and [21]. Moreover, Adom [22] suggested that residential prosumers generally exhibit a higher degree of price tolerability to electricity bills than industrial ones, implying that the responsiveness of household ESS owners in SES, as posited earlier, maybe too idealistic after being incentivized.

Individual prosumers invest in the ESS to complement the surplus power generated by their PV. The primary motivation for these prosumers to install an ESS is to enhance the flexibility of accessing and utilizing the stored energy according to their demands. However, once the SES comes into play as an aggregated system, the liberty of ESS becomes more of a luxury than a given due to the constraints enacted by the governing SES regulation. Additionally, the current SES often requires participants to perform additional actions, like optimizing rent capacity [20], reporting electricity usage information [18], submitting demands and bidding prices through auctions [23], and participating in internal energy trading [24]. It is worth questioning the attractiveness of the SES for high price-tolerance residential prosumers who experience diminished control over ESS and face increased complexity of additional actions. There is no literature study on establishing SES among these prosumers with high price-tolerances.

Prosumers with a high price-tolerance often imply that they are insensitive to cost savings in electricity bills and price incentives [25]; therefore, they do not respond to incentives provided by the ESP in the conventional SES model. Many studies employed methods aimed at improving incentive [26] or attempting to change their attitudes through education and information dissemination [27]. However, these prosumers commonly are often excluded from the construction of SES. This is because directly involving high price-tolerance prosumers in constructing an SES may not be suitable. Their unregulated charging and discharging behaviors could introduce uncertainty to the SES. Moreover, their reluctance to engage in additional actions may limit the potential profitability of the SES. Therefore, it is essential to propose a new model for the high price-tolerance prosumers, which requires no further actions and prioritizes freedom storage as a prerequisite for SES construction. Additionally, implementing a matching mechanism becomes necessary to identify suitable prosumers and ensure profitability for all participants.

Faced with frequent fluctuations in electricity market prices, participants in the electricity market need to adjust their trading strategies to achieve maximum returns. In recent years, DRL algorithms offered a new opportunity to tackle such challenges. The deep deterministic policy gradient (DDPG) algorithm and the twin delay deep deterministic policy gradient (TD3) algorithm were employed [28-30] to solve pricing optimization and coordination strategy problems. However, when considering the high price-tolerance prosumers, ensuring profits without changing their electricity consumption behaviors or requiring additional actions becomes more challenging. Moreover, compared to the conventional optimization goals on trading strategies, there is a need to optimize additional coefficients to achieve the objective of matching ESP and prosumers. Therefore, enhancing the training and optimization efficiency of the DRL algorithm is urgent.

As identified above, the following knowledge gaps should be addressed: 1) Many residential prosumers typically have a high price tolerance and may be unwilling to cooperate with an ESP to change their electricity consumption behaviors or carry out additional actions; 2) There is a lack of an effective mechanism to identify and select suitable high price-tolerance prosumers to participate in the SES; 3) To bridge the above gaps, efficient framework, method and algorithm need to be proposed. Addressing these limitations and gaps, our main contributions are as follows:

1) A framework is proposed to aggregate an SES among high price-tolerance residential prosumers, which includes construction and operation steps. This framework allows prosumers to participate without requiring additional actions or coordination with ESP.
2) A credit-based electricity deposit and withdrawal service (DWS) method is devised, where prosumers have the freedom to deposit or withdraw electricity with dynamic coefficients and receive electricity gains. A matching mechanism is proposed to facilitate mutual selections between the ESP and prosumers, ensuring that prosumer electricity bills are reduced as expected. The ESP employs this mechanism to screen users to maximize overall profitability.
3) A combined neighboring experience pool replay (CNEPR)-TD3 is proposed, which can search for similar training sets within a certain distance using multiple labels and reconstruct the experience pool. The modified TD3 can provide reasonable dynamic coefficients and energy trading strategies, effectively improving the overall profit of SES.

## II. SES Aggregation Framework with High Price-Tolerance Prosumers

### A. High Price-Tolerance Prosumers

Residential prosumers commonly have a high tolerance for electricity bills [22] for several reasons:

**Limited consumption and benefit:** Many residential prosumers' monthly electricity expenses are not significant, with the electricity bill representing only a small portion of their living expenses. While participating in the novel energy business model can lead to some savings on electricity bills, the returns are less significant in the total electricity bills.

**Limited capability:** Residential prosumers have limited means to control their electricity usage and may not have the capability for optimization operations, which may not necessarily bring profits.

Residential prosumers who have a high tolerance level for electricity bills. This implies that incentivizing them to modify their electricity consumption behaviors with low rewards may be impractical. Therefore, when faced with high price-tolerance prosumers, their electricity consumption behaviors can be seen as an inherent attribute that is not easily changed. Due to their uncontrolled charging and discharging behavior often introduces uncertainty and reduces the profitability of the SES, high price-tolerance prosumers are often excluded from the construction of SES. Therefore, we design a new SES framework specifically intended for the construction of SES with high price-tolerance prosumers.

### B. SES Framework-Construction and Operation

The framework, as shown in Fig. 1, consists of a short-term construction step and a long-term operation step.

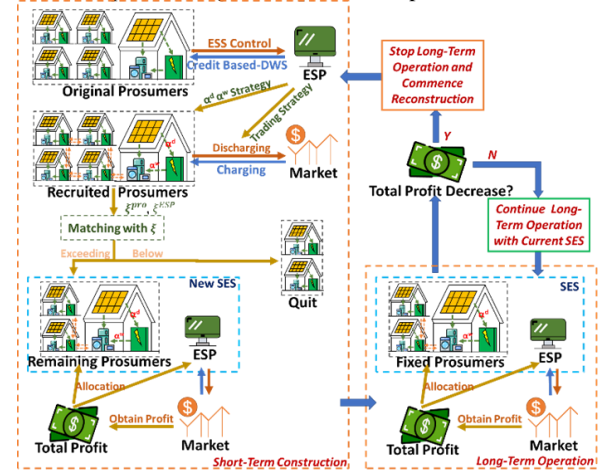

Fig. 1 Framework for SES construction and operation among price-tolerance prosumers

**SES Construction:** Before the construction of the SES, there are an ESP and a certain number of prosumers, referred to as original prosumers. Then, the ESP recruits PV-ESS prosumers willing to participate in the SES for a short-term construction, referred to as recruited prosumers in Fig. 1. Although the ESP gains control over the ESS to construct the SES, it is essential for the SES to prioritize meeting households' electricity deposit and withdrawal needs through DWS (to be detailed in Section III). Therefore, the ESP needs to optimize the dynamic coefficients of DWS and optimize the SES's charging and discharging strategies to participate in the wholesale electricity market. In addition, the prosumers typically expect to reduce their electricity bills after participating in the SES. However, to seek maximum profitability, the ESP often screens the prosumers, proposing a matching mechanism (to be detailed in Section IV) between the ESP and the prosumers for mutual selections. With the matching mechanism, some prosumers may choose to exit the SES due to a mismatch between their electricity consumption patterns and the dynamic coefficients provided by the ESP, resulting in lower profits for them. The remaining prosumers and the ESP continue the long-term operation of the SES.

**SES Operation:** After a short-term construction, the number of prosumers is fixed, referred as fixed prosumers. The overall SES can generate additional profits through participating in energy trading. This portion of the profits will be allocated among the ESP and all remaining prosumers. It is noted that during the long-term operation, it is necessary to monitor the overall profit of the SES. If the overall profit falls below 80% of the profit during the construction period, it is necessary to stop the long-term operation and start reconstruction. Otherwise, the current SES can continue to operate in the long term.

It's important to note that the short-term construction timescale is seven days, which allows for testing profitability under various electricity price scenarios and electricity consumption patterns. The long-term operational timescale is not fixed and ends only when profits drop below 80%, at which point the long-term operation is terminated, and reconstruction begins. Regarding the proposed framework, several key issues need to be addressed: the reasonable dynamic coefficients for DWS, the matching mechanism between the ESP and the prosumers, and the optimization of the trading strategy and dynamic coefficient strategy. It is important to note that the framework is not only limited to the prosumers with high price tolerance but also applies to the easily incentivized prosumers. Through different DWS coefficients, easily incentivized prosumers will change their electricity consumption behaviors, making achieving the predetermined profit goals easier and becoming SES participants.

### III. CREDIT-BASED DEPOSIT AND WITHDRAWAL SERVICE

#### A. Credit-based DWS Model

Credit is the concept introduced in the shared energy model to record prosumers' changes in charging and discharging [18]. However, the existing credit-based model cannot clearly measure whether prosumers' behaviors impact the SES trading and scheduling. Moreover, although from the prosumer's perspective, it is possible to directly charge and discharge the SES, in fact, the ESP may trade this part of the electricity directly in the electricity market without physical charging and discharging (to be detailed in Section III.D Virtual Storage and Extraction). To avoid ambiguity, we refer to the prosumers' charging operation as deposit and the discharging operation as withdraw. Therefore, with the concept of credit-based energy sharing, a credit-based DWS model is proposed as follows:

$$\Gamma_{j,t+1} = \Gamma_{j,t} + \left(\alpha_t^{dps} * P_{j,t}^{dps} * \eta^{adm} - \alpha_t^{wtd} * P_{j,t}^{wtd}\right) * \Delta t \quad (1)$$

where $\Gamma_{j,t+1}$ represents the credit points of prosumer $j$ at time $t$; $P_{j,t}^{dps}$ represents the electric power deposited by prosumer $j$, indicating the charging process for SES; $P_{j,t}^{wtd}$ represents the electric power withdrawn by prosumer $j$, indicating the discharging process for SES; $\alpha_t^{dps}$ represents the dynamic deposit coefficient, $\alpha_t^{wtd}$ represents the dynamic withdrawal coefficient; from the prosumer side, $\eta^{adm}$ can be considered as the administration cost of SES during the deposit and withdrawal process for prosumers, but in reality, it's merely the loss associated with charging and discharging, denoted as $\eta^{adm}=\eta^{ch}/\eta^{dis}$, where $\eta^{ch}$ and $\eta^{dis}$ are the charging and discharging efficiency, respectively. With the proposed DWS, the ESP aims to ensure that the electricity loss during the deposit and withdrawal process is smaller than that when prosumers control the ESS themselves, which means that $\alpha_t^{dps} \geq 1$ and $\alpha_t^{wtd} \leq 1$. Indeed, the introduction of (1) is not primarily aimed at incentivizing prosumers to change their electricity consumption behaviors, especially for those with high price tolerance who are less likely to modify their consumption patterns. Instead, its purpose is to provide rewards to prosumers whose consumption behaviors align with the dispatching and trading needs of the SES.

#### B. Factors for Dynamic Coefficients

The deposit and withdrawal coefficients $\alpha_t^{dps}$ and $\alpha_t^{wtd}$ have three main factors: the supply-demand ratio of the SES, the storage state of the SES and the potential for future profitability. The supply-demand ratio $F_t^{SDR*}$ is proposed to describe the supply-demand status of all prosumers in the SES, which can be expressed as:

$$F_t^{SDR*} = \sum_{j=1}^{J} \frac{P_{j,t}^{dps} - P_{j,t}^{wtd}}{P_{MAX}} \quad (2)$$

where $P_{MAX}$ represents the max charging and discharging power for the SES. The second factor $F_t^{SOC*}$ is related to the current storage state of the SES, as follows:

$$F_t^{SOC*} = \frac{E_t}{E_{MAX}} \quad (3)$$

where $E_t$ and $E_{MAX}$ represent the storage state at time t and max capacity of SES, respectively. The third factor, named arbitrage opportunity $F_t^{AO*}$, indicates the potential for future profitability, can be expressed as:

$$F_t^{AO*} = \lambda_{t+1}^{spot,Pred} - \lambda_t^{spot,Real} \quad (4)$$

where $\lambda_{t+1}^{spot,Pred}$ and $\lambda_t^{spot,Real}$ represent the predicted spot price at time $t+1$ and the actual spot price at time $t$ in the electricity market, respectively. $F_t^{SDR*}$, $F_t^{SOC*}$ and $F_t^{AO*}$ need to be normalized to the range [0, 1], denoting as $F_t^{SDR}$, $F_t^{SOC}$ and $F_t^{AO}$, respectively.

#### C. Balance for Deposit and Withdrawal

We propose DWS with dynamic coefficients that allow prosumers to deposit and withdraw electricity through the SES similarly to their own ESS, without requiring additional actions. There are often cases of excessive deposits or withdrawals by the prosumers, which makes it challenging to ensure zero credit within the same day. To address this issue, we provide the following settlement method:

$$R^{CR} = \sum_{t=1}^{T} \sum_{j=1}^{J} (-\Gamma_{j,t} * \lambda^{wtd}), if\ \Gamma_{j,t} < 0 \quad (5)$$

$$C^{CR} = \sum_{t=1}^{T} \sum_{j=1}^{J} (\Gamma_{j,t} * \lambda^{dps}), if\ \Gamma_{j,t} > 0 \quad (6)$$

where $R^{CR}$ and $C^{CR}$ represent the unified settlement of unsettled revenue and cost at the end of the trading day. Positive credit indicates that the prosumer has excess electricity that is not withdrawn. This surplus is purchased from the prosumer at the price $\lambda^{dps}$, which is considered as the cost for the SES. On the other hand, negative credit indicates that the prosumer withdraws more electricity than the deposit. This surplus electricity is sold to the prosumer at a price represented by $\lambda^{wtd}$, which is considered as the profit for the SES. For the prosumer, these two components result from improper planning and impose an additional burden on the SES. Therefore, we set $\lambda^{dps}$ and $\lambda^{wtd}$ to be $\beta^{FIT} * \lambda^{FIT}$ and $\beta^{TOU} * \lambda^{TOU}$, with $\beta^{FIT} < 1$ and $\beta^{TOU} > 1$.

#### D. Virtual Storage and Extraction

From the prosumer's perspective, any $P_{j,t}^{dps}$ is considered as charging the SES. However, due to storage constraints or pricing considerations, there are some situations where the electricity deposited by prosumers cannot be charged to the SES. Instead, it is directly sold on the wholesale market. This situation is referred to as virtual storage, and it encompasses three scenarios: 1) when the net power exceeds a certain range, $F_t^{SDR} > \beta^{SDR,UP}$, 2) when the battery SOH exceeds a certain range, $F_t^{SOC} > \beta^{SOC,UP}$, and 3) future arbitrage opportunity is at a large level, $F_t^{AO} > \beta^{AO,UP}$. The revenue for virtual storage can be expressed as:

$$R^{VS} = \sum_{t=1}^{T} \sum_{j=1}^{J} P_{j,t}^{dps,V} * \lambda_t^{SPO} * \Delta t \quad (7)$$

where $P_{j,t}^{dps,V}$ denotes the virtual storage power by prosumer $j$. On the contrary, when 1) the net power falls below a certain range, $F_t^{SDR} < \beta^{SDR,LOW}$, 2) the battery SOH falls below a certain range, $F_t^{SOC} < \beta^{SOC,LOW}$, and 3) electricity prices are at future arbitrage opportunity is at a low level, $F_t^{AO} < \beta^{AO,LOW}$, virtual extraction is implemented, and the cost can be expressed as:

$$C^{VS} = \sum_{t=1}^{T} \sum_{j=1}^{J} -P_{j,t}^{wtd,V} * \lambda_t^{SPO} * \Delta t \quad (8)$$

where $P_{j,t}^{wtd,V}$ denotes the virtual extraction by prosumer $j$.

### IV. MATCHING MECHANISM BETWEEN ESP AND PROSUMERS

Prosumers can participate in constructing the SES and obtain some benefits without changing their electricity consumption and storage patterns. They still expect to achieve the desired bill reduction threshold through DWS. As the ESP aims to maximize profits, it cannot aggregate all prosumers and should select participants who match their scheduling and trading strategy. To address this, based on the credit-based DWS, we propose a matching mechanism for mutual selections between the ESP and the prosumers.

*A. Matching Mechanism-based Contract*

For prosumers, DWS is an electricity gain service that can effectively reduce their bills. Prosumers want to maximize their benefits from the SES with DWS, and there is a lower threshold, assumed as $\xi^{pro}$, below which they do not participate, and above which they decide to participate in the SES. It should be noted that the $\xi^{pro}$ is an attribute of the prosumer group, which is determined by many factors of the prosumer group, including age, income, electricity bill level, education level, and even community groups' attitudes towards renewable energy [31, 32].

On the other hand, for ESP, the DWS helps identify prosumers whose electricity consumption behaviors match the SES scheduling strategy, either based on their existing behaviors or behaviors influenced by DWS incentives. When DWS can effectively help a prosumer reduce the bill, it means that the prosumer has a high level of matching with ESP's scheduling strategy. Aggregating these prosumers helps to increase the overall profit of the SES. Therefore, the lower threshold for the matching level of aggregated prosumers is denoted as $\xi^{ESP}$, and the final threshold $\xi$ needs to be selected as the maximum between $\xi^{pro}$ and $\xi^{ESP}$, expressed as:

$$\xi = \max\{\xi^{pro}, \xi^{ESP}\} \quad (9)$$

With $\xi$ as the contract threshold, the matching mechanism-based contract between ESP and prosumers can be expressed as follows:

$$\mathcal{R}_{bill,j}/\mathcal{B}_j^{ORI} \geq \xi \quad (10)$$
$$\mathcal{R}_{bill,j} = \mathcal{B}_j^{ORI} - \mathcal{B}_j \quad (10.a)$$
$$\mathcal{B}_j^{ORI} = \sum_{t^b} \lambda_{t^b}^{TOU} * (D_{t^b}^i - G_{t^b}^i) * \Delta t \quad (10.b)$$
$$t^b \in \{t | G_{j,t} \leq D_{j,t}, E_t^{ESS,j} \leq 0\} \quad (10.c)$$

where $\mathcal{R}_{bill,j}$ represents the household electricity bill reduction for prosumer $j$, which can be expressed as (10.a); $\mathcal{B}_j^{ORI}$ represents the household electricity bill for prosumer $j$ before participating in the SES, can be expressed as (10.b), and $\mathcal{B}_j$ represents that after participating in the SES, as detailed in Section V; In (10.b), before aggregating the SES, prosumer $j$ needs to pay the electricity bill for time $t^b$, as (10.c), when their demand is greater than PV generation, and their household ESS does not have electricity available for usage; $\xi$ represents the contract condition, only when this condition is satisfied, prosumer $j$ continues to participate in the SES; otherwise, prosumer $j$ should quit the SES.

*B. Conventional Profit Allocation*

The bill reduction $\mathcal{R}_{bill,j}$ that prosumer $j$ can effectively achieve can be considered as a reward for their highly matching electricity consumption behaviors. In addition to bill reduction, ESP can further generate profits by controlling the SES participants in the electricity wholesale market, which can be denoted as $\mathcal{R}_{SES}$ (to be detailed in Section V). Since the SES units are invested by the prosumers and operated by the ESP, the final profit $\mathcal{R}_{SES}$ needs to be further allocated among two parts: ESP and prosumers. We introduce the Sharply Value for profit allocation between ESP, denoted by $\mathcal{R}_{ESP}^{SV}$, and all prosumers, denoted by $\mathcal{R}_{pro}^{SV}$. Defining $X/\{i\}$ to be the set of all parts after removing part $i$, the marginal contribution of part $i$ to a coalition $S$, $S \subseteq X/\{i\}$, is

$$\epsilon_i(S) = v(S \cup \{i\}) - v(S) \quad (11)$$

where $v(S)$ represents the total payoffs that the members of $S$ can obtain by cooperation. The SV for each part in our model can be calculated as follows:

$$\phi_i = \sum_{S \subseteq X/\{i\}} \frac{|S|!(n-|S|-1)!}{n!} \epsilon_i(S) \quad (12)$$

where $n$ represents the total number of parts and is set to 2 in our work.

With the introduction of Sharply Value, $\mathcal{R}_{ESP}^{SV}$ and $\mathcal{R}_{pro}^{SV}$ can be readily obtained. Then, we further allocate $\mathcal{R}_{pro}^{SV}$ to each prosumer by their capacities as follows:

$$\mathcal{R}_{pro,j}^{SV} = \frac{E_{MAX}^{pro,j}}{\sum_{j=1}^{J} E_{MAX}^{pro,j}} * \mathcal{R}_{pro}^{SV} \quad (13)$$

where $E_{MAX}^{pro,j}$ represents the max capacity for the j-th prosumer's household ESS.

## V. PROFIT MAXIMIZATION WITH DYNAMIC DWS COEFFICIENTS OPTIMIZATION WITH CNEPR-TD3

During the construction and operation of SES, ESP should provide DWS dynamic coefficient optimization and economic operation strategies for SES. Meanwhile, prosumers only need to maintain their original electricity consumption and charging and discharging patterns, without any additional actions. Through the matching mechanism described in Section IV, both ESP and some prosumers are satisfied with the profits, which serves as a prerequisite for the construction of SES. However, a portion of prosumers, unsatisfied with the profits, choose to exit after the short-term construction step and do not participate in the long-term operation step. Therefore, the difference between the short-term construction step and long-term operation step is merely the number of prosumers, while the DWS dynamic coefficients and operation strategies remain consistent. This section introduces the optimization of DWS coefficients and operation strategies through reinforcement learning.

*A. Revenue of SES*

The ESP maximizes profit by aggregating prosumers to participate in the electricity market through DWS. The objective function of the whole SES can be expressed as follows:

$$\mathcal{R}_{SES} = R^{TRD} + R^{CR} - C^{CR} + R^{VS} - C^{VS} - C^{SES} \quad (14)$$

where $R^{CR}$ and $C^{CR}$ represent the unified settlement of unsettled credits at the end of the trading day; $R^{VS}$ and $C^{VS}$ represent the revenue and cost for virtual storage and extraction; $R^{TRD}$ represents the trading revenue in the electricity market, expressed as:

$$R^{TRD} = \sum_{t=1}^{T} (\lambda_t^{SPO} * (P_t^{SPO,ch} * \eta^{ch} - P_t^{SPO,dis}/\eta^{dis}) * \Delta t) \quad (15)$$

where $\lambda_t^{SPO}$ represents the real electricity price in the wholesale market at time t; $P_t^{SPO,ch}$ and $P_t^{SPO,dis}$ represents the electric power purchased from the wholesale market that is charged to the SES, and the electric power discharged from the SES that is sold to the wholesale electricity market. $C^{SES}$ represents the degradation cost of all SES units, expressed as:

$$C^{SES} = \sum_{t=1}^{T} \frac{k_e}{100} \sum_{j=1}^{J} \gamma_j^{SES} \left( \begin{array}{c} P_t^{SPO,dis} \\ + \sum_{j=1}^{J} (P_{j,t}^{wtd} - P_{j,t}^{wth,V}) \end{array} \right) * \Delta t \quad (16)$$

where $k_e$ represents the slope of the linear approximation of the battery life as a function of the cycles, and $\gamma_j^{SES}$ the unit capital investment on the batteries for prosumer $j$ [33, 34].

The energy storage state of SES at time t+1 $E_{t+1}^{SES}$, its constraint and the charging and discharging power constraints are respectively given as:

$$E_{t+1}^{SES} - E_t^{SES} = \sum_{j=1}^{J} \left( (P_{j,t}^{dps} - P_{j,t}^{dps,V}) * \eta^{ch} - (P_{j,t}^{wtd} - P_{j,t}^{wtd,V})/\eta^{dis} \right) * \Delta t + (P_t^{SPO,ch} * \eta^{ch} - P_t^{SPO,dis}/\eta^{dis}) * \Delta t (17)$$
$$0 \leq E_t^{ESS} \leq E_{MAX}^{SES} \quad (18)$$
$$0 \leq \sum_{j=1}^{J} (P_{j,t}^{dps} - P_{j,t}^{dps,V}) + P_t^{SPO,ch} \leq P_{MAX}^{SES,ch} \quad (19)$$
$$0 \leq \sum_{j=1}^{J} (P_{j,t}^{wtd} - P_{j,t}^{wtd,V}) + P_t^{SPO,dis} \leq P_{MAX}^{SES,dis} \quad (20)$$

where $E_{MAX}^{SES}$ represent the maximum capacity of the SES; $P_{MAX}^{SES,ch}$ and $P_{MAX}^{SES,dis}$ represent the maximum charging and discharging power of the SES.

*B. Bill of Prosumer*

With the DWS provided by ESP, prosumer $j$ can deposit excess electricity in SES and withdraw it when there is a demand. The electric power for storing and exacting can be described as follows:

$$P_{j,t}^{stg} = G_{j,t} - D_{j,t}, if\ G_{j,t} > D_{j,t} \quad (21)$$
$$P_{j,t}^{ext} = D_{j,t} - G_{j,t}, if\ G_{j,t} < D_{j,t}\ and\ \Gamma_{j,t} > 0 \quad (22)$$

where $D_{j,t}$ and $G_{j,t}$ represent the electricity demand and PV generation for prosumer $j$ at time t.

Despite prosumers transferring control of ESS to ESP, they can still maintain their original ESS storage patterns through DWS services. Consequently, they can automatically deposit excess PV-generated electricity when it exceeds their household demand and withdraw it when their household demand exceeds the PV generation and their credit $\Gamma_{j,t}$ is positive. Since their credit $\Gamma_{j,t}$ is updated every hour, there is a possibility that the credit may become negative.

According to the DWS with dynamic coefficients $\alpha_t^{dps}$ and $\alpha_t^{wtd}$, the electricity bill for prosumer $j$ can be represented as follows:
$$\mathcal{B}_j = C_j^{NET} - R_j^{CR} + C_j^{CR} \quad (23)$$
where $C_j^{NET}$ represents the cost of net load of prosumer $j$, expressed as:
$$C_j^{NET} = \sum_{t=1}^T (\lambda_t^{TOU} * P_{j,t}^{net} * \Delta t) \quad (24)$$
$R_j^{CR}$ and $C_j^{CR}$ represent the revenue and cost from settling credits at the end of each day, expressed as:
$$R_j^{CR} = \sum_{t=1}^T (\Gamma_{j,t} * \lambda^{dps}) \quad (25)$$
$$C_j^{CR} = \sum_{t=1}^T (\Gamma_{j,t} * \lambda^{wtd}) \quad (26)$$

### C. Optimization Objective

The final optimization objective function includes two parts: the overall profit of the SES and the reduction in the prosumer's bills, which can be expressed as follows:
$$max\ Obj = \mathcal{R}_{SES} + \sum_{j=1}^J \mathcal{R}_{bill,j} \quad (27)$$

For our proposed DWS-based SES construction and operation, providing a larger $\alpha_t^{dps}$ or a smaller $\alpha_t^{wtd}$ to prosumers can effectively enhance the profit of prosumers, thereby attracting a more significant number of prosumers. However, this undoubtedly affects the total profit obtained from the participation of SES in the electricity market. Conversely, prioritizing the profits from the SES's participation in the electricity market too much will impact the reduction in the prosumers' bills, thereby affecting their enthusiasm for prosumers and potentially leading to a decrease in the number of participants. Therefore, the objective function presented in (27) considers both the profit of SES and the reduction of the prosumer's bills. Due to the introduction of the matching mechanism in our work, the objective function becomes non-convex and non-continuous, making it potentially unsolvable by conventional optimization algorithms [21, 35]. Therefore, we use a DRL algorithm to simultaneously optimize the DWS strategy and the trading strategy, aiming to maximize the final objective function.

### D. Optimization of Dynamic DWS Coefficient and Trading Strategy with CNEPR-TD3

The ESP needs to provide reasonable dynamic coefficients to construct the SES with DWS for prosumers and then schedule the SES to participate in energy trading in the wholesale market. This process can be modeled as a Markov decision process (MDP) [36], which can be characterized by a tuple $<\mathcal{S}, \mathcal{A}, \mathcal{P}, \mathcal{R}>$, where $\mathcal{S}, \mathcal{A}, \mathcal{P}$ and $\mathcal{R}$ stand for the state, action, transition function and reward, respectively. The DWS coefficient and trading problem can be reformulated as the following MDP, and the agent (ESP) aims to optimize its policy to get the maximum return in the whole trajectory. In our work, the state is the feedback from the environment based on the agent's action and previous state, which includes the supply-demand ratio $F_t^{SDR}$, storage state of the SES $F_t^{SOC}$, and potential arbitrage opportunity $F_t^{AO}$. The action contains DWS coefficients $\alpha_t^{dps}$ and $\alpha_t^{wtd}$, and exchanged power $P_t^{SPO,ch}$ and $P_t^{SPO,dis}$. The reward for the agent ESP is (27). A CNEPR modified TD3 is proposed to provide optimized dynamic and exchanged power in the electricity market.

The conventional TD3 algorithm incorporates experience replay to reduce sample correlation and enhance sample utilization efficiency. However, it is essential to recognize that not all empirical samples have an equal impact on the training process. Randomly selecting samples with uniform probability may lead to including low-value or even insignificant samples in the training process. This could potentially result in suboptimal outcomes during TD3 training. Therefore, the multiple experience pool replay (MEPR) strategy [37] and multi-level experience pool replay (MLEPR) strategy [38] are proposed to select prioritized samples for updating networks in TD3.

To address the specific challenge of our proposed task, it is crucial to create multiple individual experience replay pools based on neighboring labels, which encompass the supply-demand ratio $F_t^{SDR}$, storage state of the SES $F_t^{SOC}$, and potential arbitrage opportunity $F_t^{AO}$. Therefore, a CNEPR technique is proposed to update the networks based on the priority of empirical samples, giving higher importance to neighboring and high-value samples. Fig. 2 illustrates the CNEPR technique.

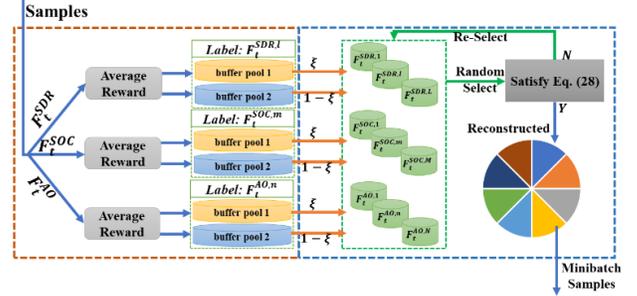

Fig. 2 Combined neighboring experience pool replay technique

The steps for constructing and utilizing multiple independent buffer pools through the CNEPR technique are as follows:

1) During the network initialization process, $F_t^{SDR}$, $F_t^{SOC}$ and $F_t^{AO}$ are divided equally into L, M and N sub-labels based on their ranges, respectively. Two experience pools are established separately for each sub-label to store high- and low-value samples. There are a total of (L+M+N)*2 pools, with labels denoted as:
$<F_1^{SDR,1}, F_2^{SDR,1}, \cdots F_1^{SDR,l}, F_2^{SDR,l}, \cdots F_1^{SDR,L}, F_2^{SDR,L}>$, and
$<F_1^{SOC,1}, F_2^{SOC,1}, \cdots F_1^{SOC,m}, F_2^{SOC,m}, \cdots F_1^{SOC,M}, F_2^{SOC,M}>$, and
$<F_1^{AO,1}, F_2^{AO,1}, \cdots F_1^{AO,n}, F_2^{AO,n}, \cdots F_1^{AO,N}, F_2^{AO,N}>$.
where subscript 1 is used to store high-value samples, and subscript 2 to store low-value samples. The average reward of coupled pools with the same superscript, such as $F_1^{SDR,l}$ and $F_2^{SDR,l}$, is set to 0.

2) When a new sample is added with labels $F_k^{SDR,l}$, $F_k^{SOC,m}$ or $F_k^{AO,n}$, (k=1,2), first, obtain the rewards corresponding to the superscripts that match the three labels. Then calculate the reward of the sample itself. If the reward is greater than the corresponding reward, store it in the corresponding pool with subscript 1. Otherwise, store it in pool 2. Repeat this process three times corresponding to three rewards, respectively, and then renew the reward.

3) During the training, samples are randomly selected from pools 1 and 2 under the same label with probabilities $\xi$ and $1 - \xi$, respectively, and are combined into a new experience buffer pool. Therefore, there are a total number of (L+M+N) pools, as green pools in Fig. 2.

4) Then, the real label $F_k^{SDR,l'}$, $F_k^{SOC,m'}$ and $F_k^{AO,n'}$, (k=1,2), is obtained corresponding to the current training data, and the randomly selected pools with labels $F_k^{SDR,l}$, $F_k^{SOC,m}$ and $F_k^{AO,n}$, (k=1,2), from the new combined pools. If the randomly selected pools satisfy
$$|l' - l| + |m' - m| + |n' - n| \leq \tau \quad (28)$$
where $\tau$ represents a distance, they can be used to reconstruct the new training pool; Otherwise, reselect and repeat step 4).

5) Finally, the mini-batch samples are randomly selected from the reconstructed pool.

## VI. CASE STUDY

### A. Experiment Setting

In case studies, the DWS service operates 24 hours a day, with the DW coefficients updated every hour. The SES construction has a trial period of 7 days, after which it transitions to long-term operation once the construction is complete. During operation, the average profit of the SES is monitored. If the average profit remains below 80% of the profit during the setup period for a continuous period of 7 days, a reconstruction is initiated. Participants are re-matched and reselected for the next operating cycle. Table I lists the parameter settings for intraday balance and virtual storage extraction.

TABLE I PARAMETER SETTING FOR SES BALANCE AND VIRTUAL OPERATION

| $\beta^{FIT}$ | $\beta^{TOU}$ | $\beta^{SDR,UP}$ | $\beta^{SDR,LOW}$ | $\beta^{SOC,UP}$ | $\beta^{SOC,LOW}$ | $\beta^{AO,UP}$ | $\beta^{AO,LOW}$ |
|---|---|---|---|---|---|---|---|
| 0.8 | 1.2 | 0.9 | 0.1 | 0.9 | 0.1 | 0.85 | 0.15 |

TABLE II Parameter setting for household PV-ESS

| PV | Battery | | | |
|---|---|---|---|---|
| $P^m$ | $E^m$ | $\eta^{ch}$ | $\eta^{dch}$ | $P^m$ |
| 10kW | 15kWh | 0.95 | 1.05 | 7.5kW |

This study is assumed to contain 2000 high price-tolerance prosumers who have installed household PV and ESS and are potential SES participants. Table II lists the parameters for PV and battery. The PV home electricity data are obtained from the Ausgrid data, and the energy market electricity prices are from the Australian Energy Market Operator (AEMO) from July 2021 to June 2022.

Before providing dynamic DW coefficients, ESP should calculate the potential arbitrage opportunity $F_t^{AO}$, which involves the price prediction method. We choose the conditional time series generative adversarial network (CTSGAN) as the electricity price prediction method. The hyperparameter settings of CNEPR-TD3 are given as follows: the critic learning rate and actor learning rate are set to 0.001, respectively; the discount factor is set to 0.9; the numbers of sub-labels for three labels are set to 5; the selection probability for experience pool 1 is set to 0.8; the distance $\tau$ is set to 0.6; the capacities of two experience pools are set to 100000; the update interval of the strategy network is set to 2; the maximum number of episodes is set to 10000. All simulations are implemented in Python 3.7 with the packages of Pytorch on VSCode and run on a PC with an AMD RYZEN 9 3950X CPU and an NVIDIA RTX 3080 GPU.

In subsequent experiments (Sections B, C and D), Case 1 is used to verify the proposed CNEPR-TD3 algorithm, contract threshold $\xi$ and virtual storage and extraction trigger condition selection. Afterwards, we employ Cases 2, 3, and 4 for comparison with Case 1, aiming to identify the contributions of our proposed DWS with dynamic coefficients and matching mechanism to the proposed model through ablation experiments in Section VI.E.

**Case 1:** The model uses a matching mechanism and credit-based DWS (w/ dynamic coefficients, w/ virtual operation). Dynamic coefficients and the scheduling strategy of the SES need to be optimized.
**Case 2:** The model uses a matching mechanism and credit-based DWS (w/ dynamic coefficients, w/o virtual operation).
**Case 3:** The model uses a matching mechanism and credit-based DWS (w/ dynamic coefficients, w/ virtual operation). However, the re-matching process is discarded, which means that the SES will continue to run in its initial state indefinitely. **Case 4:** The model uses credit-based DWS (w/ fixed coefficients, w/ virtual operation). The credit is solely used for recording the deposit and withdrawal of electricity [20], and no matching between ESP and prosumers is conducted. Only the scheduling strategy of the SES needs to be optimized.

Finally, with Case 1, the relationship between dynamic DW coefficients and three related factors is analyzed in Section VI.F.

*B. Performance of CNEPR-TD3*

The DW coefficients $\alpha_t^{stg}$ and $\alpha_t^{ext}$ and trading strategies $P_t^{SPO,ch}$ and $P_t^{SPO,dis}$ simultaneously in our proposed model (Case 1) to test the effectiveness of the proposed CNEPR-TD3 algorithm. Two modified experience-pool-based TD3 [37, 38], the conventional TD3 and DDPG are introduced as benchmarks. The comparison results are presented in Fig. 3.

The vertical axis represents the reward. It can be observed that all TD3 algorithms exhibit superior convergence performance, while the conventional DDPG algorithm exhibits poor convergence, characterized by large oscillations at the end of the learning process. The MLEPR strategy, which utilizes a multi-level experience pool, slightly improves performance compared to basic TD3; however, it enables TD3 to converge faster but does not provide further improvements in the later stages. The convergence effect of MEPR and MLEPR is similar before 6000 iterations, but because MEPR can recognize high- and low-value samples, its performance surpasses MLEPR's after 6000 iterations. Based on the mechanism of recognizing high- and low-value samples in MEPR, the proposed CNEPR-TD3 can utilizes $F_t^{SDR}$, $F_t^{SOC}$ and $F_t^{AO}$ labels to find the neighboring training set through pool reconstruction, demonstrating relatively superior performance in the early stages of training. Compared to MEPR-TD3 and MLEPR-TD3, CNEPR-TD3 can significantly improve the reward by more than A$500 per day.

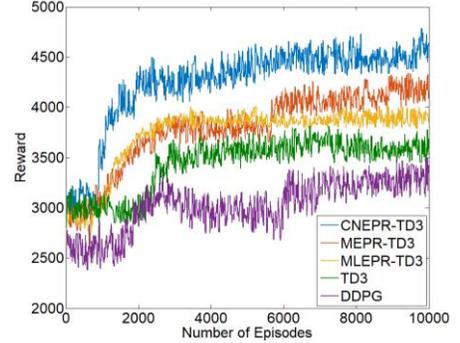

Fig. 3 Comparison of reward with different DRL algorithms

TABLE III Profit comparison with different DRL methods for Jul. 2021 and Jan. 2022 (A$/day)

| | | Bill Reduction | Participants' Profit | ESP's Profit |
|---|---|---|---|---|
| Jul. 2021 | CNEPR-TD3 | 1.56 | 0.72 | 1086.48 |
| | MEPR-TD3 | 1.57 | 0.70 | 1046.09 |
| | MLEPR-TD3 | 1.52 | 0.64 | 953.57 |
| | TD3 | 1.73 | 0.56 | 735.28 |
| | DDPG | 1.84 | 0.49 | 642.39 |
| Jan. 2022 | CNEPR-TD3 | 1.67 | 0.85 | 1435.60 |
| | MEPR-TD3 | 1.64 | 0.80 | 1257.65 |
| | MLEPR-TD3 | 1.68 | 0.75 | 1167.42 |
| | TD3 | 1.79 | 0.70 | 949.94 |
| | DDPG | 1.91 | 0.59 | 785.07 |

Table III lists the average bill reduction, participants' allocated profit, ESP's profit with the proposed CNEPR-TD3 and other benchmarking algorithms for July 2021 and January 2022. One can find that all algorithms can provide dynamic coefficients that meet participants' expectations, effectively reducing the bill. However, when comparing the three improved TD3, the dynamic coefficients provided by DDPG and basic TD3 result in excessive bill reductions for participants, with allocated profit obtained by participants of A$1.73/day and A$1.84/day in July 2021, respectively. For the SES, excessively high bill reductions indicate unreasonable dynamic coefficients. Moreover, it also leads to suboptimal optimization of market trading strategies, thereby affecting the overall profit. As a result, both individual profit allocation and ESP profit allocation are significantly reduced, with ESP's profit of A$642.39 (DDPG) and A$735.28 (TD3). All modified TD3 can effectively provide reasonable coefficients. Therefore, for participants, the daily bill reduction amounts are similar. However, our proposed CNEPR-TD3, which utilizes a pool reconstruction method based on neighboring experience replay, can provide better trading strategies. This results in daily ESP profit of A$1086.48 (July 2021) and A$1435.60 (January 2022), while MEPR-TD3 only achieves profits of A$1046.09 (July 2021) and A$1257.65 (January 2022), respectively.

Based on the above comparative analysis, it can be observed that our proposed CNEPR-TD3 algorithm exhibits superiority for trading strategy and dynamic coefficient optimization of DWS. Subsequent optimizations related to dynamic coefficients and trading strategies are conducted with CNEPR-TD3.

*C. Comparison of Different $\xi^{pro}$ for Threshold $\xi$ Selection*

Comparison of profit for ESP and prosumers with different $\xi^{pro}$ are simulated and analyzed to select the best contract threshold $\xi$ with Case 1. ESP needs to conduct a detailed investigation and set an appropriate value for $\xi^{pro}$ to ensure that prosumers are willing to accept DWS and participate in the SES. It should be noted that the $\xi^{pro}$ is an attribute of the prosumer group, which is determined by many factors of the prosumers [31]. For ESP, selecting the appropriate value for $\xi^{ESP}$ can help maximize the profit of the SES. According to (9), integrating $\xi^{pro}$ and $\xi^{ESP}$ can obtain the optimal value of $\xi$ for the current SES.

We test the performance of our proposed framework at different values of $\xi^{pro}$, ranging from 0% to 100% with intervals of 10%, to simulate the willingness of different prosumers. The results include the number of participants, average profits for prosumers, and corresponding profits for ESP. In addition, we propose a new index to indicate the deposit-withdrawal gain as: $G_{dw} = \frac{1}{T}\sum_{t=1}^{T} \alpha_t^{dps}/\alpha_t^{wtd}$. Since $\alpha_t^{dps} \geq 1$ and $\alpha_t^{wtd} \leq 1$, $G_{dw}$ must be larger than 1. Therefore, the SES has negligible loss. If $G_{dw} > 1/\eta^{adm}$, it indicates negative loss during the deposit-withdrawal process, which means that the withdrawal electricity exceeds the deposit amount.

TABLE IV COMPARISON OF REMAINING PROSUMER NUMBER, REDUCTION BILL AND ESP'S PROFIT WITH DIFFERENT $\xi^{pro}$

| $\xi^{pro}$ | Obj | Number | Bill Reduction | Participants' Profit | ESP's Profit | $G_{dw}$ |
|---|---|---|---|---|---|---|
| 0% | 1855.47 | 2000 | 0.13 | 0.40 | 793.33 | 1.03 |
| 10% | 2889.94 | 1823 | 0.56 | 0.51 | 935.81 | 1.13 |
| 20% | 4087.73 | 1734 | 1.04 | 0.66 | 1144.44 | 1.25 |
| 30% | 4695.30 | 1509 | 1.56 | 0.78 | 1171.99 | 1.37 |
| 40% | 4487.43 | 1193 | 2.08 | 0.84 | 1090.43 | 1.49 |
| 50% | 4131.86 | 937 | 2.58 | 0.91 | 855.79 | 1.62 |
| 60% | 2373.45 | 473 | 3.09 | 0.96 | 455.66 | 1.74 |
| 70% | 1587.18 | 278 | 3.58 | 1.06 | 295.61 | 1.85 |
| 80% | 413.64 | 65 | 4.14 | 1.11 | 72.37 | 1.99 |
| 90% | 117.83 | 17 | 4.66 | 1.14 | 19.32 | 2.11 |

From Table IV, one can find that there is a significant difference in the prosumers' bill reduction, allocated profit and ESP's profit under different assumed thresholds $\xi^{pro}$. When $\xi^{pro}$ is assumed to 10%, i.e., compared to the original electricity bill $\mathcal{B}_j^{ORI}$, using the DWS can further effectively reduce the bill by 10% (A\$0.56/day) for 1823 out of 2000 prosumers. Through the profit allocation process, all participants can still receive A\$0.51/day, and the ESP can receive A\$935.81/day. The total deposit-withdrawal gain is 1.133. While $\xi^{pro}$ is set to 90%, although the daily bill for each participant can be reduced at most by A\$4.66; only 17 out of 2000 individuals meet the threshold. Consequently, the ESP can only establish a small-scale SES to participate in the wholesale market, resulting in a small daily profit of only A\$19.32. Additionally, we can observe that when $\xi^{pro} \geq 10\%$, all the deposit-withdrawal gains $G_{dw}$ are greater than 1.11. This implies that even considering $\eta^{adm} = 0.9$ in the SES, the final withdrawal amount of electricity is guaranteed to be greater than the deposit amount.

We can also observe that, without considering $\xi^{pro}$, the objective function reaches its maximum daily value of A\$4695.30 when $\xi^{pro}$ is set to 30%. This implies that the optimal value for $\xi^{ESP}$ is 30%. At the same time, the ESP also achieves its maximum profit, which amounts to A\$1171.99/day. It should be noted that the lower limit of $\xi$ depends on both $\xi^{ESP}$ and $\xi^{pro}$ with (9). If the prosumers are willing to accept $\xi^{pro} \leq 30\%$, the final threshold $\xi$ is 30%. However, if the prosumers just accept $\xi^{pro} > 30\%$, the ESP can only choose actual $\xi^{pro}$ as the threshold. In the subsequent case studies, assuming that prosumers can accept $\xi^{pro} \leq 30\%$, then the final $\xi$ is set to 30%.

*D. Comparison of Different Trigger Conditions*

When $F_t^{SDR}$, $F_t^{SOC}$ and $F_t^{AO}$ meet the trigger conditions, ESP triggers the virtual storage and virtual extraction; therefore, ESP directly sells or purchases the electricity that participants store or extract in real-time on the electricity market rather than engaging in physical-level storage or extraction. In consideration of the rapid degradation caused by overcharging and over-discharging of the SES, we set $\beta^{SOC,LOW}$ to 0.1 and $\beta^{SOC,UP}$ to 0.9. Additionally, for simplicity, we set $\beta^{SDR,LOW} = 1 - \beta^{SOC,UP}$, and $\beta^{AO,LOW} = 1 - \beta^{AO,UP}$. Fig. 4 presents the total profit of SES under different $\beta^{SDR,LOW}$ and $\beta^{AO,LOW}$ with an interval of 0.05 from 0 to 0.30 in July 2021. From Fig. 4, one can find that different triggering conditions have a significant impact on the overall profit for the SES. When $\beta^{SDR,LOW} = 0$ and $\beta^{AO,LOW} = 0$, meaning that the virtual operation is only triggered when $F_t^{SOC} > 0.9$ or $< 0.1$, the daily total profit of the SES is approximately A\$4571. When $\beta^{SDR,LOW} = 0.3$ and $\beta^{AO,LOW} = 0$, the overall profit decreases to about A\$4462/day. However, when $\beta^{SDR,LOW} = 0$ and $\beta^{AO,LOW} = 0.3$, the overall profit further decreases and is about A\$4284/day. This indicates that compared to $\beta^{SDR,LOW}$, an unreasonable setting of $\beta^{AO,LOW}$ to an excessively large value will result in an even more significant decrease in overall profit. We also observe that when $\beta^{SDR,LOW}$ is set to 0.1 and $\beta^{AO,LOW}$ is set to 0.15, the overall profit is maximized at approximately A\$4753. To further analyze the reasons, we compiled the trigger rates under different triggering conditions, as shown in Table V.

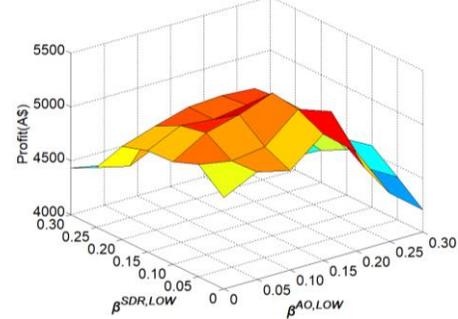

Fig. 4 Total profit of SES with different triggering conditions for virtual operation

TABLE V PROFIT, GAIN AND TRIGGER RATES WITH DIFFERENT CONDITIONS

| Triggering Condition | | | Profit(A\$) | Gain | Trigger Rate |
|---|---|---|---|---|---|
| $\beta^{SOC,LOW}$ | $\beta^{SDR,LOW}$ | $\beta^{AO,LOW}$ | | | |
| 0.1 | - | - | 4324.31 | - | 2.24% |
| | 0.1 | - | 4421.90 | 2.26% | 3.73% |
| | - | 0.15 | 4655.26 | 7.65% | 3.21% |
| | 0.1 | 0.15 | 4753.21 | 9.92% | 7.87% |
| | 0.3 | 0.3 | 4046.63 | -6.42% | 24.75% |

Table V presents the total profit, profit gain, and trigger rate under different triggering conditions. The first row of the table represents the baseline value, which corresponds to the profit when only $F_t^{SOC}$ triggers are considered. When $F_t^{SDR}$ is introduced with $\beta^{SDR,LOW}$ of 0.1 for virtual operations, the trigger rate increases from 2.24% to 3.73%, and the profit increases by 2.26%. However, if $F_t^{AO}$ is introduced instead of $F_t^{SDR}$, the trigger rate does not change significantly with 3.21%, but the profit significantly rises by 7.65%. Taking all three conditions into account, the profit increases to A\$4753.21/day, and the corresponding trigger rate is 7.87%. However, if both $\beta^{SDR,LOW}$ and $\beta^{AO,LOW}$ are set to 0.3, the trigger rate increases to 24.75%, but the profit decreases by 6.42% compared to the baseline. This demonstrates that introducing reasonable virtual operations can contribute to profit improvement, but increasing the trigger rate through unreasonable coefficient values may lead to a decrease in profit.

*E. Effectiveness Verification through Ablation Experiments*

The above case studies validate the effectiveness of the proposed CNEPR-TD3 algorithm, threshold $\xi$ selection, and triggering conditions for virtual operations. In this section, we conduct ablation experiments and calculate the profits for four cases to verify the effectiveness of the proposed model. Table VI illustrates the ESP's average daily profit, and prosumers' average daily profit, including bill reduction and allocated profit, and prosumer number from July 2021 to June 2022.

TABLE VI ABLATION EXPERIMENTS FOR PROFIT (A\$/DAY) AND PROSUMER NUMBER

| | Case | 1 | 2 | 3 | 4 |
|---|---|---|---|---|---|
| Component | Dyn. Coef. | w/ | w/ | w/ | w/o |
| | Mat. Mech. | w/ | w/ | w/ | w/o |
| | Virt. Oper. | w/ | w/o | w/ | w/ |
| | Re-Const. | w/ | w/ | w/o | w/o |
| Average Profit (A\$/day) | Prosumer | 2.43 | 2.19 | 1.63 | 0.71 |
| | ESP | 1327.92 | 1060.69 | 886.82 | 1426.68 |
| | Total | 4994.79 | 4242.76 | 3379.09 | 2853.36 |
| | Loss | - | 15.06% | 32.35% | 42.87% |
| Prosumer Number | | 1509 | 1453 | 1529 | 2000 |

Table VI shows that our proposed model (Case 1) with dynamic coefficients and the matching mechanism exhibits the best total profits,

with an average value of A$4994.79/day. In comparison to Case 1, Case 2 lacks the virtual operation component. It can be observed that both prosumers and ESPs experienced a decline in profits. The number of participants decreased by 53 compared to Case 1. Case 3 shows no re-matching and reconstruction after the profit decrease. As a result, the number of participants remains constant at 1529, the same as the initial combination. However, without reconstruction, the overall profits drop significantly. ESPs' profits are only A$886.82/day, and the prosumers' profits are A$1.63/day. In Case 4, the DWS coefficients are fixed, meaning that participants cannot obtain additional electricity gains from the SES. However, they can still receive profits from the allocation of the SES, which amounts to only A$0.71/day.

Additionally, due to the absence of a matching mechanism, it is assumed that all participants are involved. Even under this assumption, the overall profits are significantly lower than Case 1, totalling only A$2853.36/day. Based on the above analysis, it can be seen that dynamic coefficients and matching mechanisms are the most crucial components of our proposed model. Their removal would lead to a loss of 42.87%. Subsequently, re-matching and reconstruction are also important, with a loss of 32.35% when removed. Finally, removing the virtual operations results in a loss of 15.06%.

### F. Dynamic Coefficients $\alpha_t^{dps}$, $\alpha_t^{wtd}$ with Related Factors

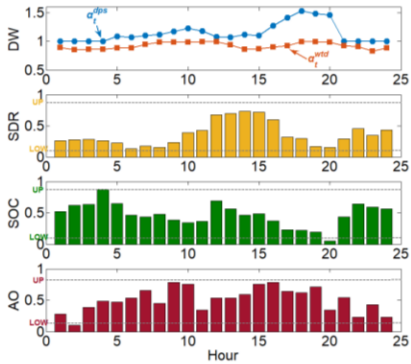

Fig. 5 Relationship between DW coefficients $\alpha_t^d$, $\alpha_t^w$ and related factors

Fig. 5 illustrates the variations in the coefficients of $\alpha_t^{dps}$ and $\alpha_t^{wtd}$ and their relationship with $F_t^{SDR}$, $F_t^{SOC}$ and $F_t^{AO}$ throughout a 24-hour period on July 30, 2022. As shown, $\alpha_t^{dps}$ reaches its highest value of 1.53 at 18:00, indicating that the depositing electricity will be credited with a coefficient of 1.53. At this time, the corresponding $\alpha_t^{wtd}$ is 1, meaning that there is no discount on electricity withdrawal during this time period. By analyzing the relationship between the coefficients $\alpha_t^{dps}$ and $\alpha_t^{wtd}$ and related factors $F_t^{SDR}$, $F_t^{SOC}$, and $F_t^{AO}$, it can be determined that the high $\alpha_t^{dps}$ value at 18:00 is due to the following reasons: 1) A low $F_t^{SDR}$ implies that a smaller portion of the PV-generated electricity can be deposited while the withdrawal of electricity for consumption increases; 2) $F_t^{SOC}$ for the SES being at a low level contributes to the higher $\alpha_t^{dps}$ value; 3) High future $F_t^{AO}$ further contributes to the high value of $\alpha_t^d$. Additionally, one can find that $F_t^{SOC}$ drops below 0.1 at 20:00, triggering virtual extraction. This means that the SES still needs to purchase electricity directly from the spot market to meet the prosumers' demand. As a result, the $\alpha_t^{dps}$ and $\alpha_t^{wtd}$ remain high.

## VII. CONCLUSION

In this work, we propose a novel SES aggregation framework with DWS and a matching mechanism, considering high price-tolerance prosumers unwilling to cooperate with ESP to change their electricity consumption behaviors or carry out additional actions. A credit-based DWS method is designed, allowing the prosumers to deposit and withdraw electricity and gain benefits. Moreover, an ESP and prosumer matching mechanism was devised to help the ESP identify prosumers whose electricity consumption behaviors align with the ESP's trading strategy, facilitating the construction of the SES. The CNEPR-TD3 is developed to optimize the dynamic coefficients in DWS and the trading strategies for participating in the wholesale electricity market, effectively improving the overall profit of the SES.

The effectiveness of the proposed CNEPR-TD3 algorithm, threshold ξ selection, and triggering conditions for virtual operation are validated by simulation results. Four cases are used for ablation experiments, and the results show that the design of dynamic DWS and the matching mechanism have the most significant impact on the overall SES profit, with an increase of 42.87%.